\documentclass[aps,twocolumn,showpacs,superscriptaddress]{revtex4}
\usepackage{graphicx}   
\usepackage{epstopdf}
\usepackage{epsfig}
\usepackage{dcolumn}
\usepackage{bm}	
\usepackage{amsmath}			
\usepackage{amsfonts}			
\usepackage{amssymb}			
\usepackage{latexsym}			
\usepackage{color}

\begin{document}

\title{Weibel instability-mediated collisionless shocks in laser-irradiated dense plasmas:\\ Prevailing role of the electrons in the turbulence generation}
\author{C. Ruyer}\email{charles.ruyer@cea.fr}
\affiliation{CEA, DAM, DIF, F-91297 Arpajon, France}

\author{L. Gremillet}\email{laurent.gremillet@cea.fr}
\affiliation{CEA, DAM, DIF, F-91297 Arpajon, France}

\author{G. Bonnaud}
\affiliation{CEA, Saclay, INSTN, F-91191 Gif-sur-Yvette, France}

\begin{abstract}
We present a particle-in-cell simulation of the generation of a collisionless turbulent shock in a dense plasma driven by an ultra-high-intensity laser pulse.
From the linear analysis, we highlight the crucial role of the laser-heated and return-current electrons in triggering a strong Weibel-like instability, giving rise to a magnetic turbulence able to isotropize the target ions.
\end{abstract}

\maketitle

\section{Introduction}

Theoretical and numerical modeling of collisionless turbulent shocks is important for understanding various high-energy astrophysical
environments, where they are held responsible for the generation of nonthermal particles and radiation  \cite{Drury_1983,Malkov_2001,Bykov_2011}.
The formation and evolution of these structures can now be simulated numerically from first principles over significant spatio-temporal scales using 
state-of-the-art particle-in-cell (PIC) codes \cite{Kato_2008,Spitkovsky_2008,Keshet_2009,Martins_2009,Nishikawa_2009,Haugbolle_2011,Sironi_Spitkovsky_2011,Stockem_2014}.
These numerical studies have demonstrated that, for initially unmagnetized colliding electron-ion flows of high enough velocities, the well-known Weibel/filamentation instability
\cite{Medvedev_1999,Achterberg_2007a,Achterberg_2007b} provides the small-scale magnetic turbulence needed for efficient dissipation of the bulk flow
energy and Fermi-type acceleration of suprathermal particles \cite{Sagdeev_1966,Lemoine_2006}. 
According to simulations, shock formation proceeds
along the following `standard' scenario: electron-driven Weibel/filamentation instabilities \cite{Bret_Gremillet_2010,Lemoine_2011b} grow and saturate
first, leaving the electrons mostly thermalized over the overlap region; for fast enough flows, an ion-driven Weibel/filamentation instability subsequently 
develops at larger scales, causing enhanced ion scattering off amplified magnetic fluctuations; the deflected ions then accumulate in the turbulent region,
until satisfying the shock hydrodynamic jump conditions \cite{Blandford_1976}.

These numerical advances go along with rapid experimental progress towards the generation of collisionless, self-magnetized shocks by intense lasers
\cite{Takabe_2008,Kuramitsu_2011, Ross_2013, Fox_2013,Huntington_2013}.  Two main configurations are currently investigated to this goal. The first one relies upon
the interaction of two counter-propagating plasma flows generated from the ablation of foil targets by  high-energy ($\sim 0.1-1\,\mathrm{MJ}$),
nanosecond-duration laser pulses \cite{Drake_2012}. Such flows are of relatively low density ($n \ll n_c$, where
$n_c \sim 10^{21}\,\mathrm{cm}^{-3}$ is the critical density of a $1\,\mu\mathrm{m}$-wavelength laser), temperature ($T_e \sim 1\,\mathrm{keV}$)
and drift velocity ($v_i \sim 1000\,\mathrm{km}\,\mathrm{s}^{-1}$).

\begin{figure}
\centerline{\includegraphics[width=0.45\textwidth]{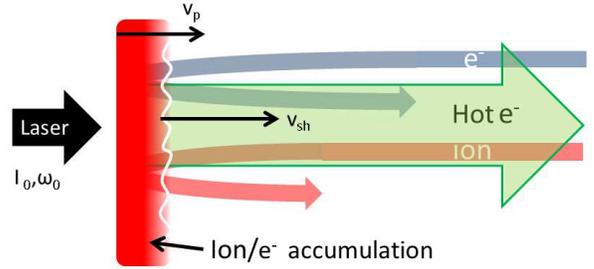}}
\caption{\label{fig:choc_laser} Sketch of the relativistic laser-driven shock formation (in the laser piston frame): electromagnetic instabilities are triggered
in the upstream by counter-streaming electron-ion flows. The magnetic fluctuations then isotropize the incoming ions,
leading to a density jump across the turbulent region fulfilling the Rankine-Hugoniot conditions.}
\end{figure}

An alternative approach, proposed by Fiuza \emph{et al.} \cite{Fiuza_2012} and further studied in this paper, hinges upon the irradiation of
an overdense plasma ($n \gg n_c$) by a relativistic-intensity ($I_0 >10^{20}\,\mathrm{Wcm}^{-2}$), picosecond-duration laser pulse
(Fig. \ref{fig:choc_laser}). During this interaction, copious amounts of electrons are heated to relativistic energies and injected
into the target. The ensuing strong charge separation and pressure gradients accelerate the surface ions to a velocity $v_i \sim 2v_p$,
where $v_p$ is the laser-driven piston (`hole boring') velocity \cite[]{Wilks_1992}. According to Ref. \cite[]{Fiuza_2012},  the various filamentation
instabilities triggered in the ion beam-plasma region may lead to the formation of a collisionless shock for intense enough lasers and/or dilute
enough targets.

In the present paper, we further investigate the physical scenario of Fiuza et al. \cite[]{Fiuza_2012}.
By means of two-dimensional PIC simulations, we will mostly focus on the laser-specific processes leading to formation of a collisionless turbulent shock, rather than on its subsequent propagation.
The numerical and physical parameters of the simulations will be presented in Sec. \ref{sec:2}.
The formation of a turbulent shock driven by a laser plane wave will be evidenced in Sec. \ref{sec:3}, and shown to fulfill the jump conditions of a strong hydrodynamic shock.
In Sec. \ref{sec:insta}, we will demonstrate the dominant role of the Weibel/filamentation instability in generating a strong electromagnetic turbulence in the upstream region.
Using exact linear theory and approximate non-linear scalings, we will assess the respective contributions of the plasma species on the instability properties and resulting saturated magnetic fluctuations.
The prevailing role of the laser-accelerated and return current  electrons will be pointed out, in contrast to the standard astrophysical scenario.
In Sec. \ref{sec:ptest}, the ion acceleration and scattering  around the shock front will be illustrated through a selection of particle trajectories.
The late time formation of strongly non-linear magnetic vortices will be discussed in Sec. \ref{sec:bubble}.
Shock generation by a focused laser wave will be considered in Sec. \ref{sec:focal}.
The final section will gather our conclusions and perspectives.
%
%

\section{Numerical setup}\label{sec:2}

The simulations have been performed using the \textsc{calder} PIC code, run in 2-D geometry. The laser pulse is modeled as an electromagnetic
plane wave linearly polarized along the $y$-axis, with a $1\,\mu\mathrm{m}$ wavelength and an intensity $I_0=3.6\times10^{21}\,\mathrm{Wcm}^{-2}$.
 This corresponds to a normalized field amplitude $A_0=eE_0/m_ec\omega_0=60$ (where $\omega_0$
is the laser frequency for a 1$\mu$m wavelength).  
The  laser intensity is held constant after a linear ramp of $30\omega_0^{-1}$ duration. 
The wave propagates along the $x>0$ direction and interacts with a fully-ionized, overdense plasma slab located at $x=40c/\omega_0$
and of maximum electron density $n_e^{(0)}$. 
The plasma ions are protons of mass $m_i/m_e=1836$ and charge $Z_i=1$. The  electron density is taken to be $n_e^{(0)}=50n_c$,
which corresponds to an electron plasma frequency of $\omega_{pe} = \sqrt{e^2n_e^{(0)}/m_e\epsilon_0}\simeq 7\omega_0$. 
These parameters are identical to those considered in Ref. \cite{Fiuza_2012}. 
The initial electron and ion temperatures are $T_{e,i}^{(0)} = 5\,\mathrm{keV}$. A $63c/\omega_0$ ($10 \,\mu\mathrm{m}$) scale-length density
ramp is added on the front surface to mimic the effect of the laser pedestal. 
The simulation grid comprises $8600\times1536$ cells with mesh
sizes $\Delta x=\Delta y=0.25c/\omega_{pe}$. The time step is taken to be
$\Delta t=0.95\Delta x/c\sqrt{2}$.  Each cell contains 50 macro-particles per species, yielding a total of $1.2\times10^9$ macro-particles.
To reduce the numerical noise, 3rd-order weight factors are used. The boundary conditions are absorbing in $x$ and periodic in $y$ for
both particles and fields.  We have checked that increasing the number of macroparticles to $100$ (with a reduced box size) or shortening
the plasma ramp does not alter significantly the results.

\begin{figure}
\includegraphics[scale=1]{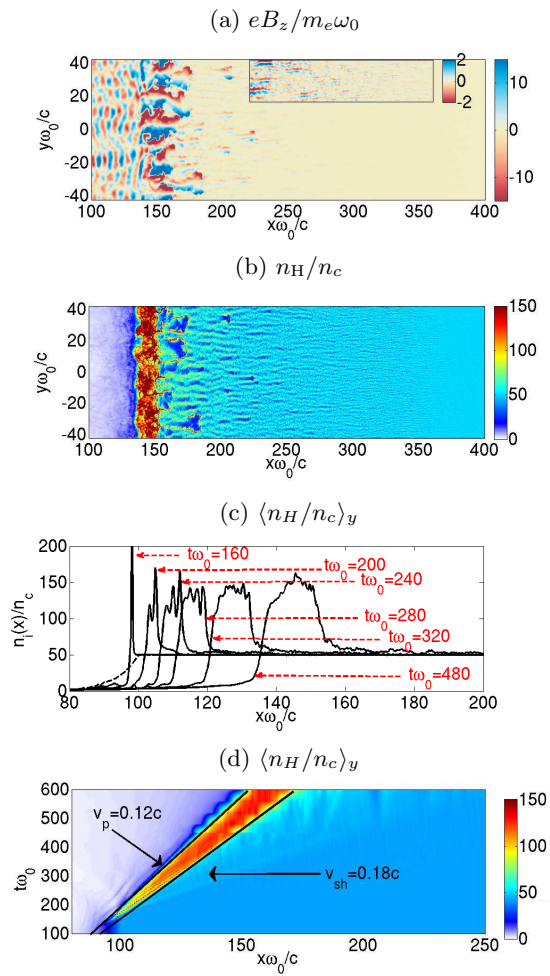}
\caption{\label{fig:nh50_a060} Shock formation in a $\mathrm{H}^+$ plasma with $A_0 = 60$ and $n_e^{(0)}=50n_c$. (a) Magnetic field $B_z$
at $t\omega_0=480$. (b) Proton density $n_\mathrm{H}$ at $t\omega_0=480$. 
(c) $y$-averaged proton density \emph{vs} $x$ at successive times. The dashed line shows the initial ion profile.
(d) $y$-averaged proton density \emph{vs} $(x,t)$.
}
\end{figure}

\begin{figure*}
\includegraphics[scale=1]{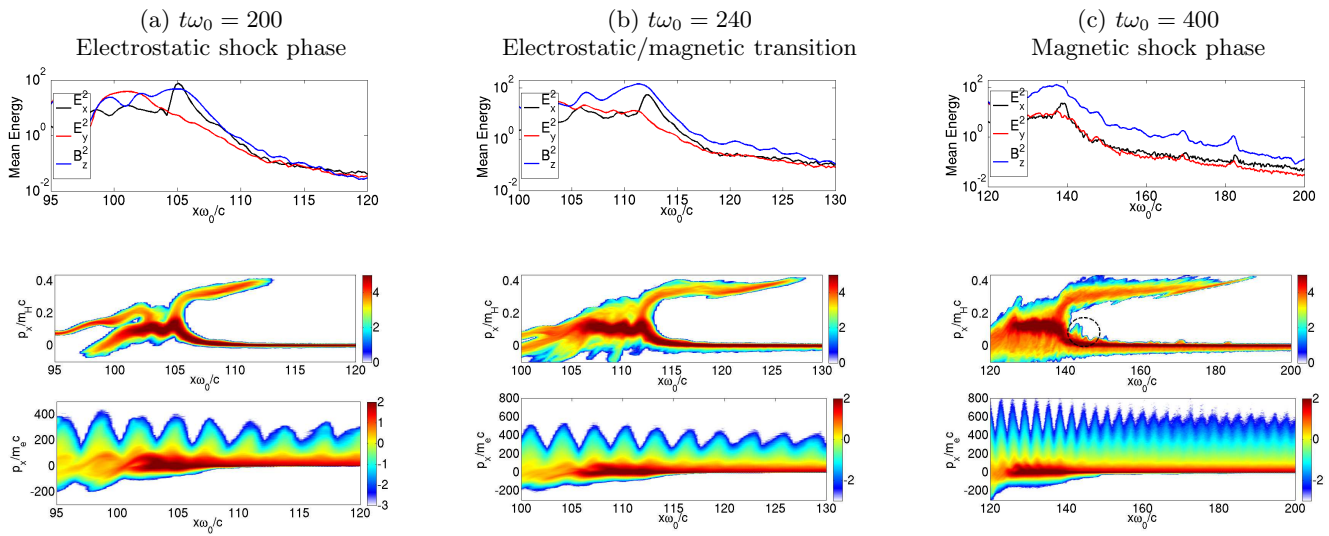}
\caption{\label{fig:PIC_transition} Shock formation in a hydrogen plasma with $A_0 = 60$ and $n_e^{(0)}=50n_c$. Upper panels: $y$-averaged
electromagnetic energies normalized to $m_ec^2n_c$. Middle and lower panels: respectively proton and electron $x-p_x$ phase spaces at (a) $t\omega_0=200$,
(b) $t\omega_0=240$ and (c) $t\omega_0=400$. The dashed contour in the ion phase space at $t\omega_0=400$
indicates the  phase space region of the magnetic vortex studied subsequently (see text).}
\end{figure*}

\section{General description}\label{sec:3}

Figure \ref{fig:nh50_a060}(a) shows that, by $t\omega_0=480$, the
Weibel/filamentation instability induced by the laser-accelerated particles flowing through the bulk plasma has given rise to strong magnetic
channels parallel to the $x$-axis. From Fig. \ref{fig:nh50_a060}(b), we see that these channels are associated to pronounced modulations in the
ion density ($\delta n_i/n_i \sim 1$).
As a result of successive coalescences and decreasing anisotropy of the particle momentum distributions
(see below), the field amplitude and filament size grow from $eB_z/m_e\omega_0 \simeq 1$ and $\lambda \simeq 3c/\omega_0$ ($\simeq 21c/\omega_{pe}$)
to $eB_z/m_e\omega_0 \simeq 10$ and $\lambda \simeq 12c/\omega_0$ ($\simeq 85c/\omega_{pe}$) when moving from $x\omega_0/c \simeq 300$ to
$x\omega_0/c \simeq 200$. Closer to the target front ($150\lesssim x\omega_0/c \lesssim 200$), the filaments exhibit kink-like oscillations, while
further growing in amplitude and size (up to $eB_z/m_e\omega_0\simeq25$ and $\lambda\omega_0/c \simeq20$) \cite{Milosavljevic_2006}. They eventually decay
into a compressed turbulent layer (further referred to as the downstream) in the interval $140\lesssim x\omega_0/c \lesssim 150$. The spatio-temporal
evolution of this shock-like structure is displayed in Figs. \ref{fig:nh50_a060}(c,d). The compression ratio relative to the unperturbed (upstream)
plasma stabilizes to a value $n_d/n_u \simeq 3$ (where $u$ and $d$ stand for upstream and downstream, respectively) by $t\omega_0 \simeq 240$,
and remains constant at later times. The right side of the compressed layer propagates at a velocity $v_\mathrm{sh}\simeq 0.18c$, while its left (irradiated)
side is pushed by the laser radiation pressure at a velocity $v_p \simeq 0.12c$. These values are consistent with a strong hydrodynamic shock induced
by the laser-driven motion of the target surface at a piston (or `hole boring') velocity \cite{Wilks_1992}
\begin{equation}
  v_p  = c\sqrt{\frac{(1+R)Z_iA_0^2}{4n_e^{(0)}m_i}}\, , \label{eq:betap}
\end{equation}
where $R$ denotes the laser reflectivity. This expression directly follows from equating the photon and ion momentum fluxes across the
laser-plasma interface (in the non-relativistic limit $v_p \ll 1$). In the present simulation, we have $R \simeq 0.4$, so that $v_p \simeq 0.12c$,
which closely agrees with the measured value. The theoretical compression ratio, $n_d/n_u$,  and velocity, $v_\mathrm{sh}$, of a nonrelativistic strong hydrodynamic
shock are
\begin{align}
  &\frac{n_d}{n_u} = \frac{\Gamma_\mathrm{ad}+1}{\Gamma_\mathrm{ad}-1}\,, \\
  &v_\mathrm{sh} = v_p\frac{\Gamma_\mathrm{ad}+1}{2}\,,
\end{align}
with $\Gamma_\mathrm{ad}$ the adiabatic index \cite{Blandford_1976}. In the present 2-D case, we have $\Gamma_\mathrm{ad}=2$, and hence
$n_d/n_u=3$ and $v_\mathrm{sh}= 0.18$, which match the simulated values. As first discussed in Ref. \cite{Fiuza_2012}, these features bear much
resemblance to those observed in simulations of Weibel-mediated electron-ion shocks in colliding cold flows \cite{Kato_2008}.

\section{Instability development and shock formation}\label{sec:insta}
 
The ion and electron $x-p_x$ phase spaces around the shock front are displayed at successive times in the middle and lower panels respectively  of
Figs. \ref{fig:PIC_transition}(a-c). The electrons accelerated in the $x>0$ direction have a large momentum dispersion ($\Delta p_x \sim 100m_ec$),
with $2\omega_0$-modulations typical of ponderomotive laser acceleration. Later on, the electron distribution broadens with time, with maximum $p_x$
increasing from $\sim 400m_ec$ at $t\omega_0 = 200$ to $\sim 800m_ec$ at $t\omega_0 = 480$.

\begin{figure}
\centerline{\includegraphics[scale=1]{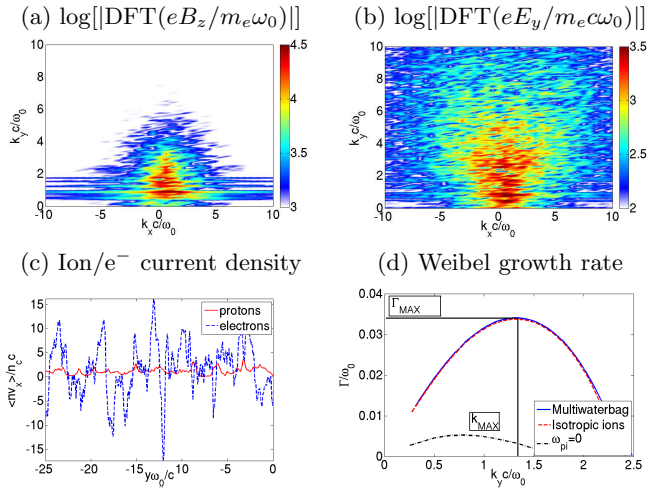}}
\caption{\label{fig:fft} Spatial Fourier transforms of $B_z$ (a) and $E_y$ (b) at $t\omega_0 = 240$ and around $x\omega_0/c=120$ (in $\log_{10}$ scale).
(c) $y$-profile of the normalized current density $\langle nv_x\rangle/n_cc$ of the electrons (blue) and protons (red). 
(d) Normalized filamentation growth rate $\Gamma/\omega_0$ solution of Eq. \eqref{eq:dispe}. 
The blue line corresponds to a multi-waterbag model of the electron and proton distributions [Figs. \ref{fig:distrib}(c,d)];  
the red line corresponds to isotropic Maxwellian protons with $T_H = 5\,\mathrm{keV}$; the black line corresponds to immobile protons.}
\end{figure}

At $t\omega_0 = 200$, partial ion reflection occurs off a shock front located at $x\omega_0/c \simeq 105$. This gives birth
to a diluted beam of density $n_{i,r}\sim0.1 n_u$, of velocities $2v_\mathrm{sh}\lesssim v_x \lesssim 0.4c$, and extending  up to $x\omega_0/c \simeq 112$. 
In the downstream region,
the ions are, on average, accelerated to the piston velocity $v_p$, while exhibiting velocity oscillations. The upper panel of Fig. \ref{fig:PIC_transition}(a)
further shows that the $y$-averaged electrostatic energy due to $E_x$ then slightly dominates the magnetic energy. All of these features suggest a transient
electrostatic, rather than magnetic, collisionless shock driven by the laser radiation pressure \cite{prl_denavit_92,Silva_2004, Zhang_2007}. At $t\omega_0 = 240$,
the reflected and upstream ion populations overlap over the space interval $112\lesssim x\omega_0/c \lesssim 128$. The peak $E_x$ energy, however, has
decreased by $50$\% in the vicinity of the shock front. Meanwhile, the magnetic energy has increased so that it prevails in both the shock foot and
downstream regions. Figures \ref{fig:fft}(a,b) plot the 2-D spatial Fourier transforms $\vert B_z(k_x,k_y)\vert$ and $\vert E_y(k_x,k_y)\vert$ in the overlap
region ($x\omega_0/c=120$) at $t\omega_0=240$. Both spectra are mostly peaked around $k_xc/\omega_0 \sim 0$ and $k_yc/\omega_0\sim 1-2$, consistently with
dominant Weibel/filamentation modes in the shock foot region.
The transition into a magnetic shock is complete by $t\omega_0 = 400$, at which time the magnetic
energy exceeds the electric energy by more than one order of magnitude. The shock front has then moved to $x\omega_0/c \simeq 140$. Note that the $E_x$ and
$E_y$ energies exhibit similar profiles in the shock foot region, except at the shock front ($x\omega_0/c \sim 140$), where the $E_x$ energy becomes larger
by a factor of $\sim2.5 $.   

\begin{figure}
\centerline{\includegraphics[scale=1]{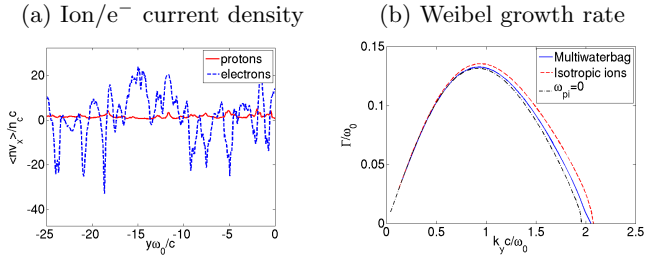}}
\caption{\label{fig:nv115} 
(a) $y$-profile of the normalized current density $\langle nv_x\rangle/n_cc$ of the electrons (blue) and protons (red) at $t\omega_0 = 240$ and $x\omega_0/c=115$. (b) Normalized filamentation
growth rate $\Gamma/\omega_0$ solution of Eq. \eqref{eq:dispe}. The blue line corresponds to a multi-waterbag model of the electron and proton
distributions;  the red line corresponds to isotropic Maxwellian protons with $T_H = 5\,\mathrm{keV}$; the black line
corresponds to immobile protons.}
\end{figure}

Figure \ref{fig:fft}(c) plots transverse lineouts of the electron and ion current density, $\langle nv_x\rangle$, at $t\omega_0=240$ in the shock foot region
($x\omega_0/c = 120$). The electron current fluctuations are seen to largely prevail over the ion current fluctuations. 
This feature holds at $x\omega_0/c = 115$ [Fig. \ref{fig:nv115}(a)] and more generally,  on both sides of the shock front until the final simulation time ($t\omega_0 = 620$).
This implies that the magnetic filaments, and the resulting shock, are mostly driven by the electrons, and not the ions.
Simulations performed with $(A_0,n_e^{(0)}/n_c)=(60,100)$, $(40,50)$, $(20,50)$ show essentially similar results.

\begin{figure}[tbp]
\centerline{\includegraphics[scale=1]{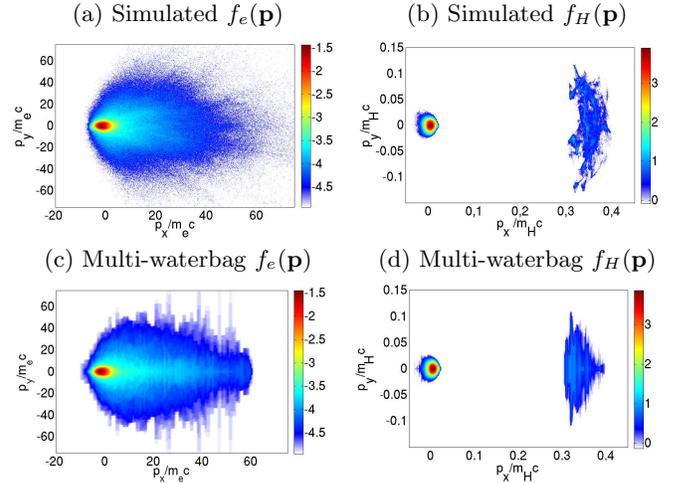}}
\caption{\label{fig:distrib} $p_x-p_y$ phase space (in $\log_{10}$ scale) of the electrons (a) and protons (b) at $x\omega_0/c=240$ and $t\omega_0=120$.
Panels (c,d) show the corresponding multi-waterbag fits used for the stability analysis (see text).}
\end{figure}

This important observation is corroborated by a linear stability analysis of the particle momentum distributions in the shock foot, displayed in Figs. \ref{fig:distrib}(a,b).
As expected \cite{Ren_2006}, the electron distribution is comprised of a diluted, high-energy tail (extending up to $\vert \mathbf{p} \vert > 100m_ec$) and of a denser
part carrying moderately-relativistic laser-accelerated electrons as well as non-relativistic return current electrons. The ion phase space exhibits two clearly-separated
structures associated to the upstream target ions (around $v= 0$) and the reflected ions (around $v_x/c \simeq 0.35$). The latter present a weakly-varying, anisotropic
($\Delta p_y \simeq 0.2m_Hc \simeq 3 \Delta p_x$) distribution, much broader than that of the upstream ions (of temperature close to the initial value of $5\,\mathrm{keV}$). 
In order to solve the dispersion relation of the Weibel/filamentation instability, it is convenient to approximate the measured distributions using a multi-waterbag
model \cite{Gremillet_2007,Bret_Gremillet_2010}
\begin{equation}
  f(\mathbf{p}) = \sum_{j=1}^N \alpha_j W_j(\mathbf{p}) \, .
\end{equation}
where $\alpha_j$ is the weight of the $j$th waterbag component.  The relativistic waterbag distributions are taken to be even functions of $p_y$ of the form
\begin{equation}
  W_j(\mathbf{p}) = \frac{1}{4P_{jx} P_{jy}} H(P_{jx} - \vert p_x- P_{jd}\vert)H(P_{jy}-\vert p_y \vert)\delta (p_z) \,,
\end{equation}
where $H$ is the Heaviside function and $P_{jd},P_{jx},P_{jy}$ are adjustable parameters. Figures \ref{fig:distrib}(c,d) display the best-fitted multi-waterbag
approximations (with $N= 10^4$) of the electron (c) and ion (d) momentum distributions. The multi-waterbag distributions reproduce the momentum fluxes
of the original distributions (a,b) to within an error of $<10 \%$.  The susceptibility tensor associated to these waterbag distributions is detailed in Appendix A
of Ref.  \cite{Bret_Gremillet_2010}.  As shown in Ref. \cite{Ruyer_Gremillet_2013}, the dispersion relation of the system can be solved for all modes
$\omega_n(\mathbf{k}) \in \mathbb{C}$ by generalizing to the relativistic electromagnetic regime the method introduced by Fried and Gould in the
non-relativistic electrostatic limit \cite{Fried_Gould_1961}. Using this procedure, the filamentation growth rate $\Gamma = \Im \omega$ is plotted as a
function of the transverse wave number $k_y$ (for $k_x = 0$) in Fig.  \ref{fig:fft}(d). 
The maximum value $\Gamma_\mathrm{max}/\omega_0 \simeq 0.035$ at $x\omega_0/c =120$
is obtained at the wave number $k_\mathrm{max}c/\omega_0 \simeq 1.3$. To assess the contribution of the ions to the instability, we have replaced their
multi-waterbag distribution by an isotropic non-drifting Maxwell-J\"uttner of temperature $T_H = 5\,\mathrm{keV}$. The resulting growth rate curve almost
coincides with that obtained with the full ion distribution, which proves that the reflected ions are not responsible for the observed filamentation instability.
This feature holds on both sides of the shock front and is illustrated in Fig. \ref{fig:nv115}(b) at $x\omega_0/c =115$.
As will be analyzed later on, because the electrons' mean energy is here comparable to the ions' (as measured in the piston frame), the electrons can
induce magnetic fluctuations strong enough to scatter the ions and entail shock formation. This contrasts with the standard scenario of Weibel-mediated
astrophysical shocks \cite{Lyubarsky_2006}, where the unstable two-stream ion distribution is considered as the key player in generating magnetic
turbulence. Although, in our case, the ion anisotropy is not the driving force behind the magnetic buildup, the thermal bulk ions, while stable \emph{per se},
may significantly enhance the electron-driven instability by mitigating space-charge effects [Fig. \ref{fig:fft}(d)]. 
This destabilizing mechanism, previously discussed in Refs. \cite{Tzoufras_Ren_2006,Ren_2006}, is demonstrated by computing the growth rate upon assuming infinite-mass ions, 
and also addressed in \cite[]{Nitin_2012}. 
As seen in Fig. \ref{fig:fft}(d),
this yields a maximum growth rate lowered by a factor $\sim 6$, and a dominant wave number down-shifted to $k_\mathrm{max}c/\omega_0 \sim 0.7$. 

\begin{figure}
\centerline{\includegraphics[scale=1]{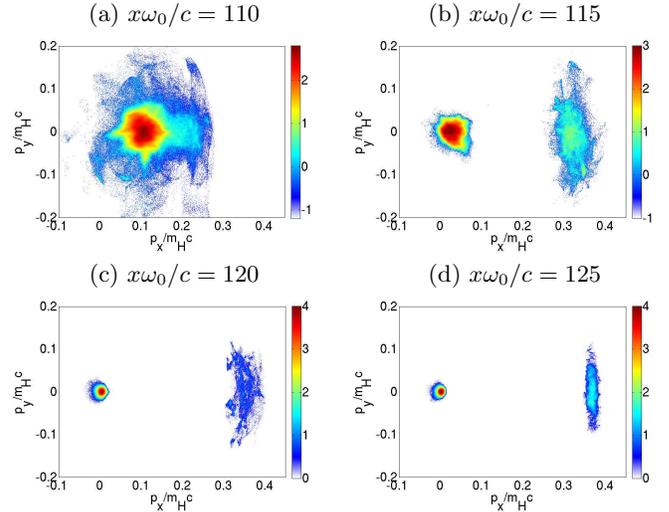}}
\caption{\label{fig:mwa} Proton $p_x-p_y$ phase space at $t\omega_0 = 240$ and various locations: $x\omega_0/c=110$ (a), $115$ (b), $120$ (c)
and $125$ (d) (in $\log_{10}$ scale).}
\end{figure}

\begin{figure}
\centerline{\includegraphics[scale=1]{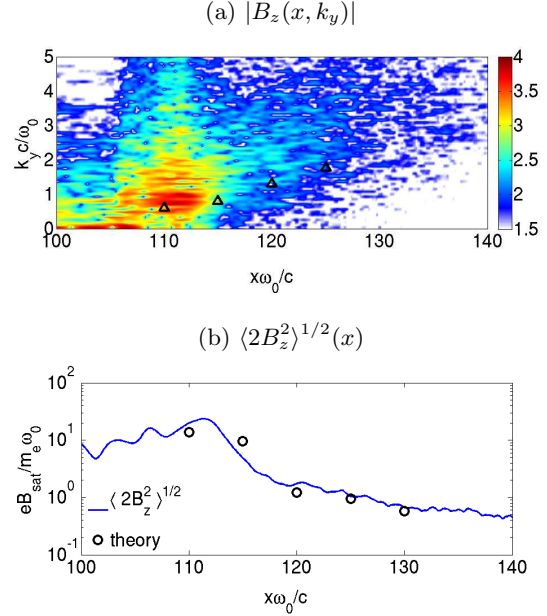}}
\caption{\label{fig:mwa_bsat} (a) Magnetic spectrum $\vert B_z(x,k_y)\vert $ (in $\log_{10}$ scale) in the upstream region at $t\omega_0=240$. The triangles
plot the fastest-growing wave numbers, $k_\mathrm{max}$, predicted from linear theory at the locations of Fig. \ref{fig:mwa}(a-d). (b) Spatial profile of the
transversely averaged magnetic field $\langle 2B_z^2 \rangle^{1/2}$ at  $t\omega_0=240$ (blue solid line). The black circles plot Eq. \eqref{eq:bsat2}, where
$k_\mathrm{sat}$ and $\gamma_{e,\mathrm{eff}}$ are measured from the simulation.}
\end{figure}

Figures \ref{fig:mwa}(a-d) plot the proton $p_x-p_y$ phase space at various locations in the upstream region at time $t\omega_0=240$. Both the bulk and reflected
proton distributions broaden as one moves closer to the shock front as a result of growing magnetic scattering. At the shock front ($x\omega_0/c = 110$), they have
merged into a relatively isotropized population [Fig. \ref{fig:mwa}(a)]. In \mbox{Fig. \ref{fig:mwa_bsat}(a)} the magnetic spectrum  $\vert B_z(x,k_y)\vert $ in the
upstream region is displayed and compared to the fastest-growing wave number predicted from linear theory using the particle distributions of Figs. \ref{fig:mwa}(a-d).
Overall, a correct agreement is obtained between linear theory and the simulated spectrum.

The spatial profile of the transversely-averaged magnetic field amplitude, $\langle 2B_z^2 \rangle^{1/2}$, at $t\omega_0 = 240$ (i.e., the approximate formation
time of the magnetic shock) is plotted in \mbox{Fig. \ref{fig:mwa_bsat}(b)}. The amplitude varies by aboyt two orders of magnitude over the upstream region
$110\lesssim x\omega_0/c \lesssim 140$. Note that at this time, the reflected ions extend to $x\omega_0/c \simeq 128$ [\mbox{Fig. \ref{fig:PIC_transition}}(b)].
At the shock front ($x\omega_0/c \simeq 112$), the magnetic field reaches a value $\langle B_z^2 \rangle^{1/2} \simeq 25m_e\omega_0/e$, comparable to the laser
field strength.

\begin{figure}
\centerline{\includegraphics[scale=1]{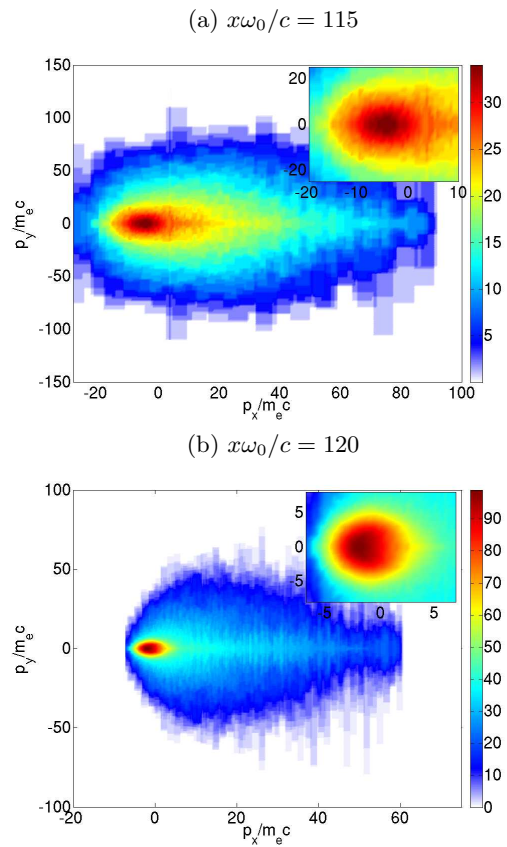}}
\caption{\label{fig:critere_weibel} Map of $\sum S_i W_i(\mathbf{p})$ at (a) $x\omega_0/c=115$ and  (b) $x\omega_0/c=120$, which measures the local contribution to the Weibel instability in the electron $p_x-p_y$ phase
space. The subpanel zooms in on the most destabilizing electrons.}
\end{figure}

In order to analyze this magnetic profile, we have assessed the effectiveness of the various parts of the electron distribution in driving the Weibel instability.
A similar evaluation was made in Ref. \cite{Ren_2006}, yet within the simplifying assumption of a purely transverse instability. Here, we adopt an alternative
approach based on our multi-waterbag model. The condition for instability of a multi-waterbag system reads $\Delta = AB -C^2 > 0$, where the factors $A$,$B$,$C$,
given in Appendix B of Ref. \cite{Bret_Gremillet_2010}, take the form of a sum over the waterbag components. For instance, we have
\begin{equation}
  C = \sum_j \omega_{pj}^2 P_{jd}/P_{jy}^2 \,.
\end{equation}
Let us now define
\begin{equation}
  C_i = \sum_{j\ne i} \omega_{pj}^2 P_{jd}/P_{jy}^2 \,,
 \end{equation}
where the index $i$ labels a given waterbag component. 
Likewise, we introduce $A_i$,  $B_i$, $\Delta_i =A_i B_i -C_i$ and $S_i=1-\Delta_i/\Delta$. The latter
expression then quantifies the stabilizing ($S_i<0$) or destabilizing ($S_i>0$) influence of the $i$th waterbag component. 
Figures \ref{fig:critere_weibel}(a,b)
display $\sum_j S_j W_j(\mathbf{p})$ using the electron distribution measured at $t\omega_0 = 240$, $x\omega_0/c = 115$ and $x\omega_0/c = 120$. 
It appears that the electrons mostly
responsible for the late-stage instability have moderate relativistic energies ($\gamma \simeq 3$ at  $x\omega_0/c = 120$ and $\gamma \simeq 8$ at  $x\omega_0/c = 115$) and, on average, negative $x$-momenta ($p_x < 0$). 
These particles are
therefore associated with the background return current induced by the higher-energy, laser-driven electrons propagating in the $x >0$ direction. 
This feature
was first pointed out in Ref. \cite{Ren_2006} under similar laser-plasma conditions. The relativistic energies attained by the return current electrons stem
from the various (electromagnetic Weibel/filamentation, electrostatic longitudinal/oblique) beam-plasma instabilities induced in the upstream region.
 
Let us now confront the simulated magnetic profile of \mbox{Fig. \ref{fig:mwa_bsat}(b)} to simple models of the saturated field accounting for the Weibel-effective
part of the electron distribution. According to the widely-used transverse-trapping model \cite{Davidson_1972,Yang_1994,Silva_2002,Kaang_2009}, saturation occurs
when the electron bounce frequency inside a magnetic filament is equal to the maximum growth rate . This yields the magnetic field amplitude
\begin{equation}\label{eq:bsat1}
  \frac{eB_\mathrm{sat}}{m_e\omega_0} \simeq \left \langle \frac{\gamma}{\beta_x} \right \rangle_\mathrm{eff}
  \left(\frac{\Gamma_\mathrm{max}}{\omega_0}\right)^2 \frac{\omega_0}{k_\mathrm{max}c} \, ,
\end{equation}
where $\langle \gamma/\beta_x \rangle_\mathrm{eff}$ denotes the average of $\gamma/\beta_x$ over the Weibel-effective electrons. We have typically
$\langle \gamma/\beta_x \rangle_\mathrm{eff} \simeq 5-10$. 
In principle, $\Gamma_\mathrm{max}$ is difficult because it is comparable to the laser frequency \cite[]{POP_Okada_2007,ArXiv_Robinson_2013}, yielding very short exponential growth and saturation time.
Using the late-time $\Gamma_\mathrm{max}$ values of Figs. \ref{fig:fft}(d) and \ref{fig:nv115}(b) therefore greatly underestimates the saturated field ($eB_\mathrm{sat}/m_e\omega_0 \simeq 0.01-0.1$),
except in the far-upstream region.
corresponds to a more strongly nonlinear regime than is assumed in the transverse-trapping model. A second estimate of $B_\mathrm{sat}$ may therefore be
derived supposing that the Weibel-effective electrons are magnetized \cite{Moiseev_1963, Lyubarsky_2006}. Equating the typical filament size,
$2\pi/k_\mathrm{sat}$ to the Larmor radius of the effective electrons, $\langle \gamma \rangle_\mathrm{eff}/eB$, gives the lower limit
\begin{equation}\label{eq:bsat2}
  \frac{eB_\mathrm{sat}}{m_e\omega_0} \simeq \langle \gamma \rangle_\mathrm{eff} \frac{k_\mathrm{sat}c}{\pi \omega_0} \, .
\end{equation}
For a numerical application, the saturated wave number $k_\mathrm{sat}$ is extracted from the spectrum of Fig. \eqref{fig:mwa_bsat}(a), while the 
Weibel-effective electrons are defined as that part of the electron phase space satisfying $\sum_j S_j W_j(\mathbf{p})> \frac{1}{2}\max_\mathbf{p} (\sum_j S_j W_j)$.
The resulting $B_\mathrm{sat}$ values are plotted at different locations as black circles in Fig. \ref{fig:mwa_bsat}(b), where they are found to capture to a
good accuracy the simulated magnetic profile. 

In summary, a perturbative analysis of the local plasma distribution functions shows that the magnetic turbulence in the upstream region mostly results from the
interplay between  the laser-accelerated and background electrons. More precisely, the dominant destabilizing effect comes from the moderate-energy return current
electrons. While the reflected ions have a negligible influence, the background ions strengthen the instability by weakening its inhibiting, electrostatic component
\cite{Tzoufras_Ren_2006,Ren_2006}. Although the typical wave numbers compare satisfactorily with linear theory, the magnetic field profile in the shock-foot region
is indicative of the magnetization of the Weibel-effective part of the electron distribution.

\section{Proton trajectories}\label{sec:ptest}

\begin{figure}[t]
\centerline{\includegraphics[scale=1]{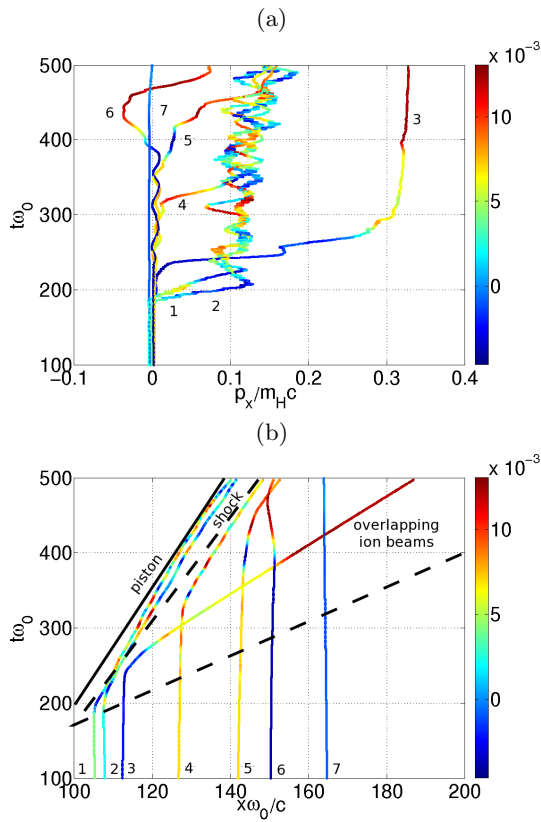}}
\caption{\label{fig:part_test}Typical proton trajectories for various initial $x$-locations: (a) $p_x(t)$ and (b) $x(t)$. The color of each curve is indexed by
$p_y(t)/m_Hc$. In (b) are also plotted the trajectories of the laser-driven piston (black solid line), of the shock front (black dashed line) and of the reflected
proton front (black dotted-dashed line). }
\end{figure}

To gain insight into the dynamics of the upstream protons, we plot in \mbox{Figs. \ref{fig:part_test}(a,b)} seven typical proton trajectories $(x(t),p_x(t))$,
originating from increasing target depths. The color of each trajectory (labeled by the particle number) is indexed by the instantaneous value of the normalized
$y$-momentum, $p_y(t)/m_Hc$. In \mbox{Fig. \ref{fig:part_test}(b)}, the trajectories of the laser-driven piston, of the shock front and of the reflected ion
front are also plotted.

Particles 1 and 2 are rapidly (over the time interval $190\lesssim t\omega_0 \lesssim 220$) accelerated forward by the electrostatic field set up by the electron
pressure gradient at the (then mainly electrostatic) shock front.  After reaching the downstream (piston) velocity ($v_x/c \simeq 0.12$), they remain confined in
the downstream region where they experience an increasing level of electromagnetic turbulence. The latter makes them oscillate in both $v_x$ and $v_y$, with
similar amplitudes $\Delta v_x \sim \Delta v_y/c \sim 0.02$. Particle 3 is more strongly accelerated by the electrostatic shock potential:  it attains a velocity
$v_x/c \simeq 0.3$, which corresponds to reflection in the shock frame.  During its main acceleration phase ($220 \lesssim t\omega_0 \lesssim 300$), its $y$-velocity
hardly varies due to a weak magnetic turbulence in the shock foot region. Later on, however, the magnetic fluctuations get strong enough to induce velocity variations
$\Delta v_y/c \sim  0.02$.

Particles 4-6 exemplify the ion dynamics in the magnetic shock regime.  Because the upstream electromagnetic turbulence has grown in amplitude and spatial extent,
they undergo an increasing number of oscillations in $v_x$ and $v_y$ while being, on average, accelerated along $x$ by the electrostatic field. The effective range
of the turbulence can be assessed by noting that particle 7, initially located at $x\omega_0/c \simeq 165$, does not exhibit any significant acceleration up to
$t\omega_0 = 500$, at which time the shock front has moved to $x\omega_0/c \simeq 145-150$. 

\section{Late-time evolution: magnetic vortices}\label{sec:bubble}

\begin{figure*}[t]
\centerline{\includegraphics[scale=1]{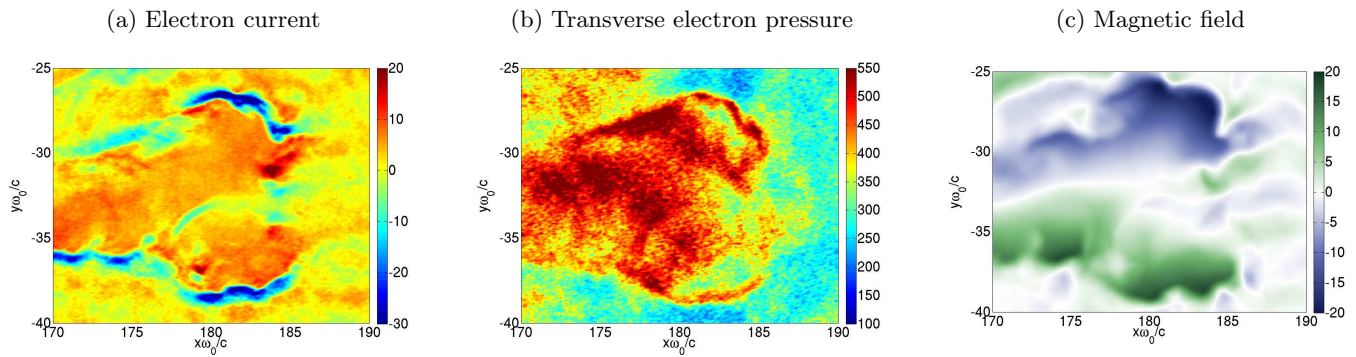}}
\caption{\label{fig:bubble} Magnetic vortex: (a) electron current density $\langle nv_x \rangle/n_cc$, (b) $y$-component of electron pressure tensor
$\langle n v_yp_y \rangle/n_cm_ec^2$ and (c) magnetic field $eB_z/m_e\omega_{0}$.}
\end{figure*}

The late-time evolution of the shock foot region shows the formation of partially depleted ion `bubbles'.
Figures \ref{fig:nh50_a060}(a-c) zoom in on the bubble located around $x\omega_0/c=180$ and $y\omega_0/c=-32$, displaying the associated electron current, transverse electron pressure and magnetic field at $t\omega_0=480$.
The bubble consists of a magnetic dipole sustained by a current of hot electrons flowing in the forward direction, neutralized by thin return-current layers at its lower and upper boundaries. 
The mean hot electron current density is $j_{e,x} \sim -6en_c c$ over a width $l_y\sim 7c/\omega_0$ , yielding a magneto-static field $B\sim  \mu_0l_y j_{e,x}/2 \sim 20 m_e\omega_0/e$.
The thickness of the return current layers is of the order of the relativistic skin depth of the upstream plasma $\sim \sqrt{\gamma}c/\omega_{pe}\sim0.5c/\omega_0$.
Similar magnetic structures have been observed in Refs.  \cite[]{ppr_kutznetov_2001,ppr_kutznetov_2005,prl_nakamura_2010} in the case of lower-density interaction regimes.
In the present case, the magnetic vortices result  from the non-linear evolution of the Weibel instability in the shock foot region, when the magnetic pressure inside the hot electron's filaments 
($ B^2/2\mu_0\sim 200 n_cm_ec^2 $) is strong enough to expel the upstream ions.
Figure \ref{fig:nh50_a060}(b) shows that the magnetic pressure is comparable to, yet lower than the hot electron pressure ($ \sim 500 n_cm_ec^2 $).
As in Refs. \cite[]{ppr_kutznetov_2001,ppr_kutznetov_2005,prl_nakamura_2010}, the vortex expands at a velocity close to the  Alfv\'en velocity, $v_a\sim B/\sqrt{\mu_0n_im_i} \sim 0.1c$.
The localized velocity jump indicated by a dashed circle in Fig. \ref{fig:PIC_transition}(c) corresponds to a few ions being reflected off a magnetic vortex. 
These magnetic structures develop significantly after the ions have become isotropised  so that the shock formation should not be affected. 
A few magnetic vortices at the early stage of their development are also visible in Ref. \cite[]{Fiuza_2012}.
Note that these structures, observed in our 2D simulation, could be absent in a 3D geometry.

\section{Influence of a finite laser spot-size}\label{sec:focal}

In order to assess whether the above scenario still holds in a realistic setting, we have run an additional simulation with a   $16\mu$m (FWHM) Gaussian laser profile.
The simulation grid has been enlarged up to $16800\times9216$ cells with unchanged  discretization.
The number of macroparticles per mesh has been reduced to $20$, so that their total number in the simulation is $6.2 \times 10^9$.
The other laser and plasma parameters are the same as before.

\begin{figure}[h]
\centerline{\includegraphics[scale=1]{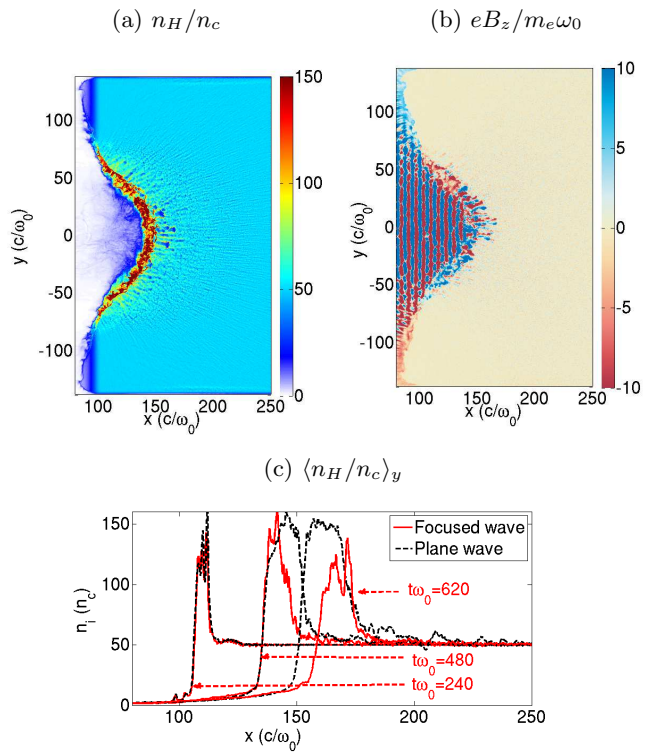}}
\caption{\label{fig:focal} (a) Proton density, (b) magnetic field at $t\omega_{0}=480$ and (c) ion density profiles averaged over the transverse direction in the region $\vert y\omega_0/c \vert<10$ (red solid line)
for a Weibel-mediated collisionless shock induced by a  laser of intensity $I_0\simeq 1.8\, 10^{21}$W.cm$^{-2}$ and $16\mu$m spot size.
The dashed black lines plot the $y$-averaged density  from the plane wave simulation.
 }
\end{figure}

The simulated ion density and magnetic field at $t\omega_0=480$ are displayed in Figs. \ref{fig:focal}(a,b) and compared with the plane-wave results in Fig. \ref{fig:focal}(c).
Because of the Gaussian shape of the laser beam, the piston velocity peaks on the laser axis ($y=0$) and decreases away from it,
resulting in a curved laser-plasma interface.
However, the proton density  of Fig. \ref{fig:focal}(a) shows a relatively flat transverse profile (independent of $y$), close to the laser axis ($\vert y\omega_{0}/c \vert<20$).
As further shown by the density lineouts of Fig. \ref{fig:focal}(c), the plasma is compressed by a factor $\sim3$, consistently with the Rankine-Hugoniot conditions.
This compressed region exhibits about five magnetic filaments, which seems to be sufficient to form the shock-like structure.
Lineouts of the proton density are plotted at different times in Fig. \ref{fig:focal}(c), and compared to those from the plane wave simulation.
At $t\omega_0=240$, the two configurations yield similar profiles.
At $t\omega_0=480$, the location of the laser-piston and the compression factor are very close in the two cases, yet the downstream region is thinner (by $\sim50$\%) in the focused configuration.
This can be explained by the curvature of the laser-plasma interface [Fig. \ref{fig:nh50_a060}(a)], which allows the ions and electrons of the downstream to leak off the $y=0$ axis.
%
%
To ensure a stable front shock over a duration $\Delta t$, the laser spot size, $D$, should be larger than $v_p \Delta t$,
otherwise, the downstream will be significantly distorted by the Gaussian piston, which may hamper the shock propagation.
In the  case of a $D=16\mu$m-focal spot, the shock should be stable up to $\Delta t \lesssim D/v_p \simeq 800\omega_0^{-1}$.
The drop in the downstream density observed at $t\omega_0=620$ (corresponding to effective interaction time of $520\omega_0^{-1}$)
supports qualitatively this estimate. 
Note that making use of a super-Gaussian laser profile may yield a more stable shock front.

\section{Conclusions}

In contrast to the currently explored experimental setup \cite[]{POP_Kugland_2013, Fox_2013, Yuan_2013,Huntington_2013} which requires $10-100$kJ-class laser facilities, 
we have found that only $1-2$kJ of intense-enough laser pulses should be able to drive Weibel-mediated shocks, which confirms the results of Ref. \cite{Fiuza_2012}.
Their formation proceeds as follows.
Rapidly after the start of the irradiation, an electrostatic shock forms and propagates.
At later times, because the hot electrons have typical energy comparable to that of the ions (in the piston frame), they are able to trigger a strong magnetic turbulence in the shock front region.
The critical role of the return-current electrons in driving the Weibel-filamentation instability has been shown, making use of a multi-waterbag decomposition scheme.
Unlike the usual framework \cite[]{Moiseev_1963,ppr_sagdeev_66,Lyubarsky_2006},
the ion heating and isotropization result from an electron-driven instability (rather than ion-driven), at least over the relatively short time-scales considered in our simulation.
During shock propagation on longer time-scales, the ions could play a significant role in triggering the upstream instability. 
We have also shown that shock formation is accessible to a broad enough focused laser wave.
However, the shock is found to decay away over time scales $\gtrsim D/v_p$

Several points have yet to be clarified to ensure the experimental feasibility of Weibel-mediated laser-induced collisionless shocks. 
The influence of collisions, ionization, radiative losses as well as 3D effects could be important.
One of the most critical issue is linked to the diagnostics of such experiments.
A major drawback of such shocks is that they develop in dense plasmas over very short space ($\sim \mu$m) and time scales ($100$fs), which greatly complicates their experimental characterization unless one disposes of an intense, short-duration x-ray probe as provided by a free-electron-laser  \cite[]{Kluge_2014}.

\section*{Acknowledgments}

The authors gratefully acknowledge Arnaud Debayle, Anne Stockem and Frederico Fiuza for interesting discussions. 
The PIC simulations were performed using HPC resources at TGCC/CCRT (Grant No. 2013-052707).

\appendix

\section{Dielectric tensors}\label{ap:suscep}
Introducing $\theta$, the angle between the wavevector and the $x$-axis, 
the linearization of the Vlasov-Maxwell equations yields the well-known dispersion relation \cite[]{Bret_2008}:
\begin{align}
  &(\omega \varepsilon_{xx}-k^2 \sin^2\theta)(\omega \varepsilon_{yy}-k^2 \cos^2\theta)\nonumber\\
  &-(\omega \varepsilon_{yz}-k^2 \cos \theta \sin\theta )^2=0\, . \label{eq:dispe}
\end{align}
The dielectric tensor reads:
\begin{equation}
\boldsymbol{\epsilon}_{\alpha\beta} = \delta_{\alpha\beta}+\sum_s \frac{\omega_{pr}^2}{\omega^2} \boldsymbol{\chi}
\end{equation}
where $\chi$ is the relativistic susceptibility tensor  given in Refs. \cite[]{Bret_Gremillet_Benisti_2010, Ruyer_Gremillet_2013, Bret_Gremillet_2010} for waterbag and Maxwell-J\"uttner distribution functions.
All  the  tensor elements can  easily be recast as a function of the phase speed and the angle $\theta$.
Equation \eqref{eq:dispe} gives 
\begin{equation}
ak^4+bk^2+c=0\, , \label{eq:dispe_oblique}
\end{equation}
with
\begin{align}
a&=(\beta_\phi^2 -\sin^2\theta)(\beta_\phi^2 -\cos^2\theta) -\cos^2\theta \sin^2\theta \, , \label{eq:dispe_obliquea}\\
b&=(\sin^2\theta-\beta_\phi^2) \sum_s \omega_{ps}^2\chi_{yy}     (\cos^2\theta-\beta_\phi^2)\sum_s \omega_{ps}^2\chi_{zz} \nonumber\\
&+2\cos\theta \sin\theta\sum_s \omega_{ps}^2\chi_{yz}\, ,   \label{eq:dispe_obliqueb}\\
c&= (\sum_s \omega_{ps}^2\chi_{yy})(\sum_s \omega_{ps}^2\chi_{zz}) - (\sum_s\omega_{ps}^2 \chi_{yz})^2\, ,\label{eq:dispe_obliquec}
\end{align}
In Eqs. \eqref{eq:dispe_obliquea}, \eqref{eq:dispe_obliqueb} and \eqref{eq:dispe_obliquec}, the subscripts have been omitted on $\chi$ for the sake of clarity. 
The wavevector then verifies
\begin{align}
  k^2 &=\frac{-b(\beta_{\phi}) + \sqrt{\Delta(\beta_{\phi}})}{2a(\beta_{\phi})}\label{eq:solg1} \, ,\\
  k^2 &=\frac{-b(\beta_{\phi}) - \sqrt{\Delta(\beta_{\phi}})}{2a(\beta_{\phi})}\label{eq:solg2} \, ,
\end{align}
 with $\Delta =\sqrt{b^2 -4ac} $.
This formulation, in which $k^2 (>0)$ is a function of $\beta_\phi$ only, 
lends itself to the efficient numerical scheme introduced by Fried and Gould \cite{Fried_Gould_1961} in a non-relativistic framework and generalized recently \cite[]{Ruyer_Gremillet_2013}.
This scheme consists, first, in determining the locus of the zeroes of $\Im \mathcal{G}(\beta_\phi)$.
This can be readily performed by means of a contour plot in a finely discretized portion of the complex $\beta_\phi$ plane. 
Then, we retain those zeroes fulfilling $\Re \mathcal{G}(\beta_\phi) > 0$ and identify $k=\sqrt{\Re \mathcal{G}(\beta_\phi)}$.
Depending on the $\beta_\phi$ domain considered, this method allows us to simultaneously solve for a set of discrete electromagnetic solutions $\omega(k,\theta)$.


\end{document}